# Automating REST API Postman Test Cases Using LLM

S Deepika Sri (csdeepikasri@gmail.com),
Mohammed Aadil S (thasinaadil@gmail.com), Sanjjushri Varshini R (sanjjushrivarshini@gmail.com),
Raja CSP Raman (raja.csp@gmail.com), Gopinath Rajagopal (prgopinath100@gmail.com),
S Taranath chan (taranath6579@gmail.com)
https://github.com/tactlabs/test-case-generation

## Abstract

In the contemporary landscape of technological advancements, the automation of manual processes is crucial, compelling the demand for huge datasets to effectively train and test machines. This research paper is dedicated to the exploration and implementation of an automated approach to generate test cases specifically using Large Language Models. The methodology integrates the use of Open AI to enhance the efficiency and effectiveness of test case generation for training and evaluating Large Language Models. This formalized approach with LLMs simplifies the testing process, making it more efficient and comprehensive. Leveraging natural language understanding, LLMs can intelligently formulate test cases that cover a broad range of REST API properties, ensuring comprehensive testing. The model that is developed during the research is trained using manually collected postman test cases or instances for various Rest APIs. LLMs enhance the creation of Postman test cases by automating the generation of varied and intricate test scenarios. Postman test cases offer streamlined automation, collaboration, and dynamic data handling, providing a user-friendly and efficient approach to API testing compared to traditional test cases. Thus, the model developed not only conforms to current technological standards but also holds the promise of evolving into an idea of substantial importance in future technological advancements.

## 1 Introduction

In today's fast-paced technological landscape, automation plays a pivotal role in enhancing productivity and efficiency. With the increasing reliance on machine learning and artificial intelligence, the demand for large datasets to train and test these systems has become paramount. This research paper addresses the pressing need for automated test case generation, particularly focusing on REST API testing, through utilizing Large Language Models (LLMs). The methodology employed in this study leverages the advanced capabilities of LLMs, including Open AI, to streamline the test case generation process for large language models. By integrating natural language understanding, LLMs can intelligently formulate test cases that cover a broad spectrum of REST API properties, thereby ensuring comprehensive testing coverage.

The model developed in this research is trained using manually collected Postman test cases for various REST APIs. This training data enables the LLMs to learn and understand the intricacies of REST API testing, facilitating the automated generation of diverse and complex test scenarios. This formalized approach not only enhances the efficiency of the testing process but also ensures the generation of high-quality test cases that accurately reflect real-world usage scenarios. One of the key advantages of this approach is its user-friendly nature. By automating the generation of test cases, the burden on software developers and testers is significantly reduced, allowing them to focus on other critical tasks. Additionally, the integration of LLMs with tools like Postman further simplifies the testing process, offering a seamless experience for users. Furthermore, this research aligns with current technological standards and holds promise for future advancements in the field of automated testing methodologies. As machine learning and artificial intelligence continue to evolve, the role of LLMs in test case generation is expected to become increasingly prominent, further revolutionizing the way software systems are tested and evaluated. In conclusion, this research paper presents a novel approach to automated test case generation using LLMs, with a specific focus on REST API testing.

## 2 Literature Review

In recent years, the integration of Large Language Models (LLMs) has garnered significant attention in software engineering, promising conversational support throughout the software lifecycle. Researchers have explored various applications of LLMs in software engineering, ranging from requirements extraction to vulnerability repair [1]. However, concerns have been raised regarding the security risks associated with code generated by LLMs such as GitHub Copilot and ChatGPT. The absence of security-sensitive datasets in LLM evaluation and the lack of security-focused evaluation metrics contribute to insecure code generation. To address this, frameworks like SALLM have been proposed to systematically assess LLMs' ability to generate secure code [2]. Advancements in automated test generation tools have also been significant, particularly in unit test generation for Java programs. While existing tools excel in structural coverage and fault detection, seamless integration within Integrated Development Environments (IDEs) remains a challenge. TestSpark, a user-friendly IntelliJ IDEA plugin, bridges this gap by supporting various test generation techniques, including LLM-based approaches [3]. However, the effectiveness of LLMs in generating unit tests, particularly in strongly typed languages like Java, remains uncertain. A study investigating the performance of three LLMs in generating JUnit tests for Java classes found mixed results [4]. Automation in software testing, particularly in unit test generation, has been a topic of interest. LLMs show promise in automating unit test generation, but there's a need for empirical assessment [5]. Similarly, existing tools in Java programs often lag in incorporating the latest advancements in LLMs. Evaluating the quality of Java unit tests generated by an OpenAPI LLM reveals promising results [6]. The application of LLMs in mobile application test script generation addresses challenges in accurately reproducing scripts across diverse devices and platforms. LLMs' adaptability to varied user interfaces and app architectures contributes to enhancing software testing practices [7]. In the realm of testing RESTful APIs, RESTTESTGEN has been introduced as an innovative approach utilizing Swagger interface definitions for automatic test case generation. Its effectiveness in revealing faults across real-world REST APIs has been demonstrated [8]. Despite the strengths of LLMs in generating human-like responses, challenges such as hallucination and limited external knowledge use persist. The LLM-AUGMENTER system addresses these issues by enhancing LLMs with plug-and-play modules for external knowledge integration [9]. The rapid rise of LLM chatbots has raised concerns about user privacy and security. Existing approaches to exposing vulnerabilities in mainstream LLM chatbots have shown limited efficacy [10]. Automated test generation tools play a crucial role in software development, with recent efforts focusing on leveraging LLMs to generate unit tests. While existing studies primarily assess code coverage, there's recognition that coverage alone may not correlate strongly with bug detection effectiveness [11]. Mobile task automation presents scalability challenges, which can be addressed by leveraging LLMs. AutoDroid represents a novel approach, integrating LLMs' commonsense and app-specific knowledge through dynamic analysis [12]. These literature surveys provide insights into distinct methodologies aimed at enhancing automated testing processes, emphasizing the integration of LLMs and dynamic optimization-based search techniques. Such methodologies aim to improve the effectiveness and scalability of testing techniques in real-world software development scenarios [13]. Similarly, the second survey highlights a paper proposing a technique for automatically generating test data to test exceptions, emphasizing dynamic global optimization-based search to ensure safety-critical systems are free from anomalous behavior [14]. Both methodologies aim to enhance automated testing processes, whether through the utilization of LLMs or dynamic optimization-based search, to improve the effectiveness and scalability of testing techniques in real-world software development scenarios [15].

# 3 Proposed Solution

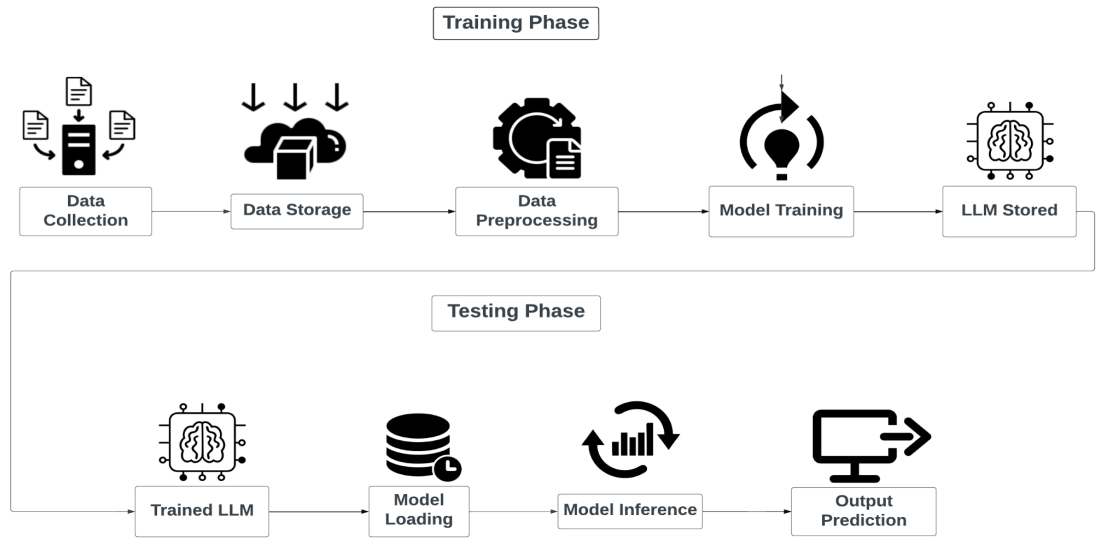

*Figure 1. Architecture Diagram of Postman Test Case Generation*

To address the challenge of automated test case generation for REST APIs using Large Language Models (LLMs), we propose a comprehensive solution that encompasses data architecture, model development, and integration into existing testing frameworks. The solution involves the following steps:

## 3.1 Training Phase

### 3.1.1 Data Collection

Data collection initiates the training phase by sourcing relevant data from diverse repositories, including databases, APIs, or files. Ensuring the comprehensiveness and representativeness of the collected data is crucial for effective model training. This step lays the foundation for the Language Model's learning process.

### 3.1.2 Data Storage

Subsequently, the collected data needs to be securely and efficiently stored for further processing. Data storage involves employing databases or data warehouses to facilitate seamless accessibility and scalability. Proper storage mechanisms ensure that the training data remains readily available for manipulation during subsequent steps.

### 3.1.3 Data Preprocessing

The collected data undergoes preprocessing to refine its quality and structure, making it suitable for training. Preprocessing tasks encompass tokenization, normalization, and potentially data augmentation. This preparatory step enhances the compatibility of the data with the model and contributes to overall performance improvement.

### 3.1.4 Model Training

The core of the training phase revolves around training the Language Model using the preprocessed data. This iterative process optimizes model parameters to minimize a predefined loss function, typically employing techniques like backpropagation and gradient descent. Through continuous adjustment, the model progressively learns to better comprehend and generate human-like text.

### 3.1.5 LLM Stored

Upon completion of training, the trained Language Model (LLM) is stored in a designated file format, often with a ".pth" extension. This file encapsulates the acquired knowledge and parameters of the model, preserving them for future utilization. Storing the LLM facilitates easy retrieval and deployment in subsequent phases or applications.

## 3.2 Testing Phase

### 3.2.1 LLM Trained

The testing phase commences with additional refinement or extension of the trained Language Model (LLM). This may involve fine-tuning the model on supplementary data or adjusting hyperparameters to enhance performance further. LLM training iteratively improves the model's accuracy and adaptability to diverse contexts and tasks.

### 3.2.2 Model Loading

Subsequently, the trained LLM is loaded into memory to facilitate inference tasks. This entails retrieving the stored model parameters from the *".pth"* file and initializing the model for utilization. Model loading ensures that the trained model is primed and ready for processing new input data.

### 3.2.3 Model Inference

With the loaded model, inference tasks can be executed on new input data. Model inference involves passing input sequences through the model to generate output predictions. Leveraging the acquired knowledge and parameters, the model interprets and generates responses or classifications based on the provided input.

### 3.2.4 Output Prediction

The culmination of the testing phase involves obtaining output predictions generated by the model during the inference process. These predictions encapsulate the outcomes of the model's processing and can be employed for various downstream tasks, such as decision-making, recommendation, or analysis. The reliability and accuracy of the output predictions are indicative of the model's effectiveness in interpreting and responding to input data.

In summary, the training phase initiates comprehensive data collection from various sources, followed by preprocessing to refine the data for training. The core step involves training the model, and optimizing parameters iteratively to minimize loss and enhance its ability to generate human-like text. Once trained, the Language Model is stored for future use. In the testing phase, the trained model undergoes additional refinement if needed before being loaded for inference tasks. Output predictions generated during inference serve as a validation of both phases, reflecting the model's capability to interpret input data accurately.

## 4 Tech Stack

### 4.1 Flask

Flask is well-suited for handling HTTP requests and serving machine learning models as APIs. Developers can define routes to accept incoming requests and use view functions to process these requests and return predictions from the machine learning models. Developers can implement their own request validation logic within Flask routes using libraries such as Flask-RESTful or Flask-Inputs.The Postman test cases are stored in a JSON file and the file is available on the Flask web page to download which can be imported into Postman.

### 4.2 Postman

Postman is used to test and debug the API endpoints. Postman is a collaboration platform for API development that allows users to design, test, and debug APIs. It is used to send HTTP requests to the Flask server to test the API endpoints and inspect and validate the responses returned by the server. It facilitates debugging and troubleshooting of API functionality. Postman is used by developers to ensure that the API endpoints are functioning correctly and returning the expected results during the development and testing phases. These additional technologies enhance the functionality and development process of your model by providing a robust API framework (Flask) for serving predictions and a comprehensive tool (Postman) for testing and debugging the API endpoints.

### 4.3 Open AI

OpenAI's language models, such as GPT-3, are implemented using deep learning techniques, particularly the Transformer architecture. These models are trained on vast amounts of text data using powerful hardware accelerators like GPUs and TPUs. Implementation of Open AI in our model typically involves fine-tuning the pre-trained model on generating test cases in postman test cases format. OpenAI provides APIs and libraries that we can use to integrate the models into their applications easily. Additionally, the deployment of these models often requires robust infrastructure to handle the computational demands and ensure efficient inference at scale.

## 5 Result

The research emphasizes the method's efficacy and inclusivity by leveraging natural language understanding to cover a wide range of API properties. The methodology relies on a formalized approach that utilizes a dataset consisting of manually collected Postman test cases to train the LLMs. This dataset serves as the foundation for creating diverse and intricate test scenarios, enabling comprehensive API testing coverage. The user-friendly nature of the approach is highlighted, reducing the workload for developers and testers while adhering to contemporary technological standards.

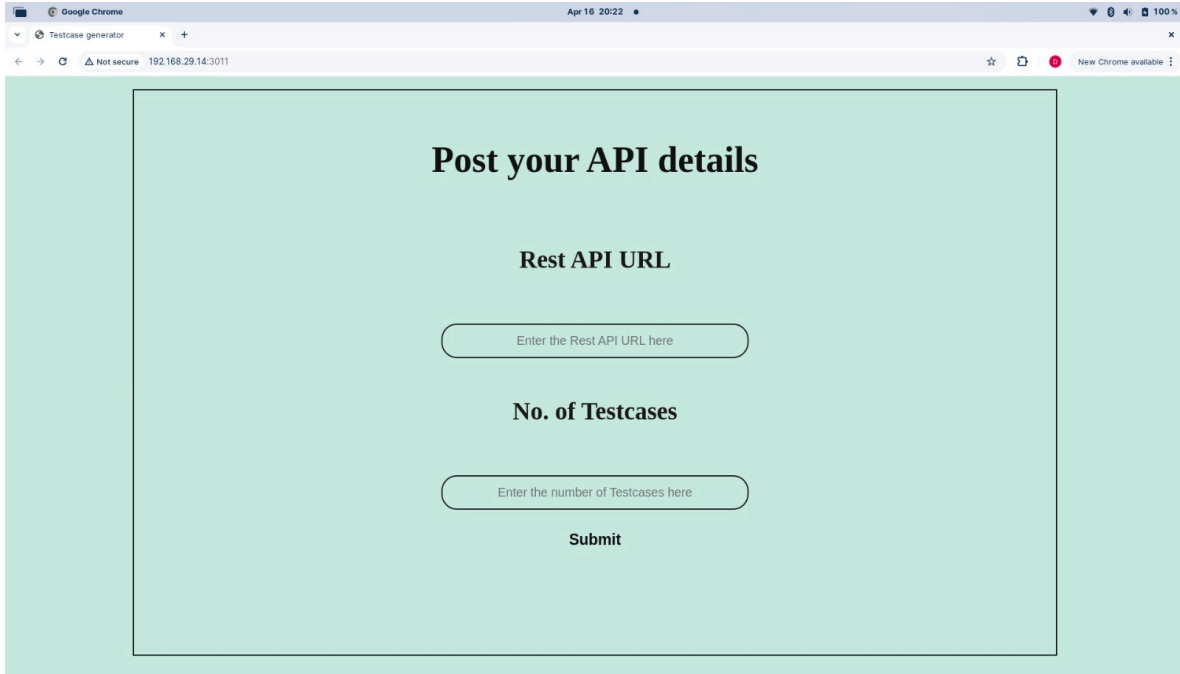

*Figure 2. Home Page*

The input to the methodology consists of just getting the REST API and number of test cases that need to be generated from the user. Figure 2 illustrates an intuitive interface, showcasing how users can input natural language descriptions of API properties, streamlining the test case generation process. This user-friendly approach, coupled with the effectiveness of OpenAI's LLMs in natural language understanding, facilitates the creation of diverse and intricate test scenarios.

*Figure 3. Output Page - Postman Test Cases Generated*

The output is a set of comprehensive test scenarios tailored specifically for testing the provided API. The approach produces intricate test scenarios that cover a wide array of REST API functionalities through a formalized methodology and utilizing manually collected Postman test cases for training. The download button in Figure 3 allows users to obtain a JSON file containing the comprehensive test scenarios, facilitating easy importation and execution within Postman for efficient testing of the REST API.

Overall, the method provides the potential of LLMs to advance automated testing practices, particularly in the context of REST API testing, by providing a more inclusive and efficient approach compared to traditional methods.

## 6 Future Scope

Over the course of test case generation research, we have utilized the OpenAI model to generate test cases. Further research can be conducted to explore several avenues for enhancing test case generation using different pre-trained models and transformers. The model can be developed to utilize the generated Postman test cases within the Postman platform without human involvement. Developing automated workflows for orchestrating the generation, execution, and evaluation of test cases can further streamline the testing process. Automated test orchestration systems can leverage LLM-generated test cases alongside other testing techniques to create end-to-end testing pipelines. Focusing on generating test cases specifically aimed at uncovering security vulnerabilities, such as injection attacks, authentication bypasses, or data leakage, is essential. Security-focused test case generation can help identify and mitigate security risks early in the software development lifecycle.

## 7 Conclusion

In conclusion, the proposed solution presents a robust framework for automating test case generation in REST API testing through the integration of Large Language Models (LLMs). By leveraging natural language understanding and a formalized methodology, the solution streamlines the process of creating comprehensive test scenarios, covering a wide range of API properties. The inclusion of manually collected Postman test cases for training enhances the model's ability to generate intricate and diverse test cases, thereby improving testing coverage and efficiency.

Furthermore, the utilization of Flask for serving the machine learning model as an API, coupled with Postman for testing and debugging API endpoints, enhances the functionality and accessibility of the solution. This tech stack not only facilitates seamless integration of the LLM-generated test cases into existing testing frameworks but also ensures robustness and reliability in the testing process. Additionally, the incorporation of OpenAI's LLMs underscores the scalability and versatility of the solution, paving the way for advancements in automated testing methodologies.

Looking ahead, future research can explore various avenues for enhancing test case generation using different pre-trained models and transformers. Further development can focus on automating workflows for orchestrating the generation, execution, and evaluation of test cases, thereby streamlining the testing process and improving overall efficiency. Overall, the proposed solution lays a solid foundation for advancing automated testing practices, particularly in the context of REST API testing, and holds promise for addressing the evolving challenges in software testing.